\documentclass[cits]{PoS}
\newcommand{\m}{M_{H^{\pm}}}
\newcommand{\g}{\,\mbox{GeV}}
\newcommand{\rg}{R_{\gamma\gamma}}
\newcommand{\rzg}{R_{Z\gamma}}
\usepackage{amsmath}
\usepackage{amssymb}
\usepackage{amsfonts}
\usepackage{times}
\usepackage{graphicx}

\title{Decay rates of the Higgs boson to two photons and\\ $Z$ plus photon in
$\mathbf{Z}_2$-symmetric Two Higgs Doublet Models}

\ShortTitle{$R_{\gamma\gamma}$ and $R_{Z\gamma}$ in the Inert Doublet Model}

\author{\speaker{Bogumi\l a~\'{S}wie\.{z}ewska}\\%
       Faculty of Physcis, University of Warsaw,\\
       Ho\.{z}a 69, 00-681 Warsaw, Poland\\
        E-mail:  \email{Bogumila.Swiezewska@fuw.edu.pl}}

\abstract{Analysis of the $125\g$ Higgs boson decay rates to $\gamma\gamma$ and $Z\gamma$ in the Inert Doublet Model is presented. We study the  constraints on the masses of the scalars (in particular the Dark Matter candidate) and their couplings to the Higgs boson, coming from the $h\to\gamma\gamma$ data and confront them with the WMAP measurements of the Dark Matter relic density.}

\FullConference{The European Physical Society Conference on High Energy Physics \\
                 18-24 July, 2013\\
                 Stockholm, Sweden}

\begin{document}
\section{Inert Doublet Model and the loop-induced Higgs boson decays}
Inert Doublet Model (IDM) is a Two Higgs Doublet Model with scalar doublets $\Phi_S$ and $\Phi_D$. IDM is exactly symmetric under the transformation $D$, such that: $\Phi_D \xrightarrow{D} -\Phi_D$, $\Phi_S \xrightarrow{D} \Phi_S$, $\Phi_{\mathrm{SM}} \xrightarrow{D} \Phi_{\mathrm{SM}}$. The particle spectrum of the IDM consist of a SM-like Higgs boson $h$ which couples to fermions and gauge bosons like the SM Higgs boson, and four dark scalars: $H$, $A$, $H^{\pm}$ which do not couple to fermions at the tree level. Due to the conservation of $D$, the lightest $D$-odd scalar is stable, providing a viable dark matter (DM) candidate. The IDM is in agreement with current theoretical and experimental constraints (LEP, LHC and WMAP).

The signal strength in the $h\to\gamma\gamma$ channel is measured at the LHC, giving the results that are consistent with the SM value ($\rg=1$) but the experimental uncertainties leave space for the new physics contributions. Unfortunately, for the $h\to Z\gamma$ channel there are still not enough data. 
In the IDM $\rg$ and $\rzg$ can deviate from 1 by two means: if the total width of the Higgs boson is augmented due to the existence of invisible decays ($h\to AA$ or $h\to HH$) or if the partial decay width of $h\to \gamma\gamma$ is modified due to the $H^{\pm}$ loop.

\begin{figure}[hb]
\centering
\includegraphics[width=.45\textwidth]{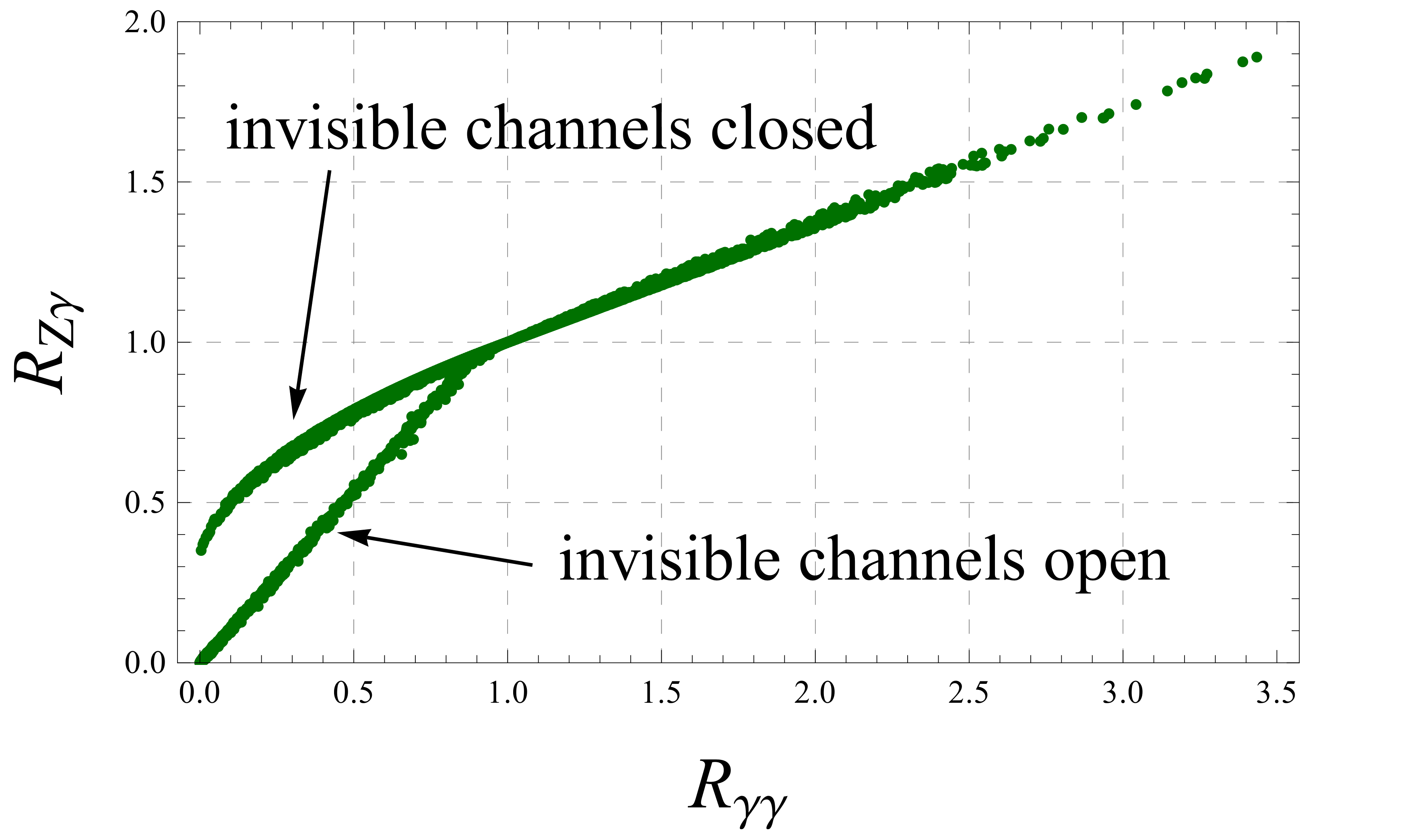}
\includegraphics[width=.42\textwidth]{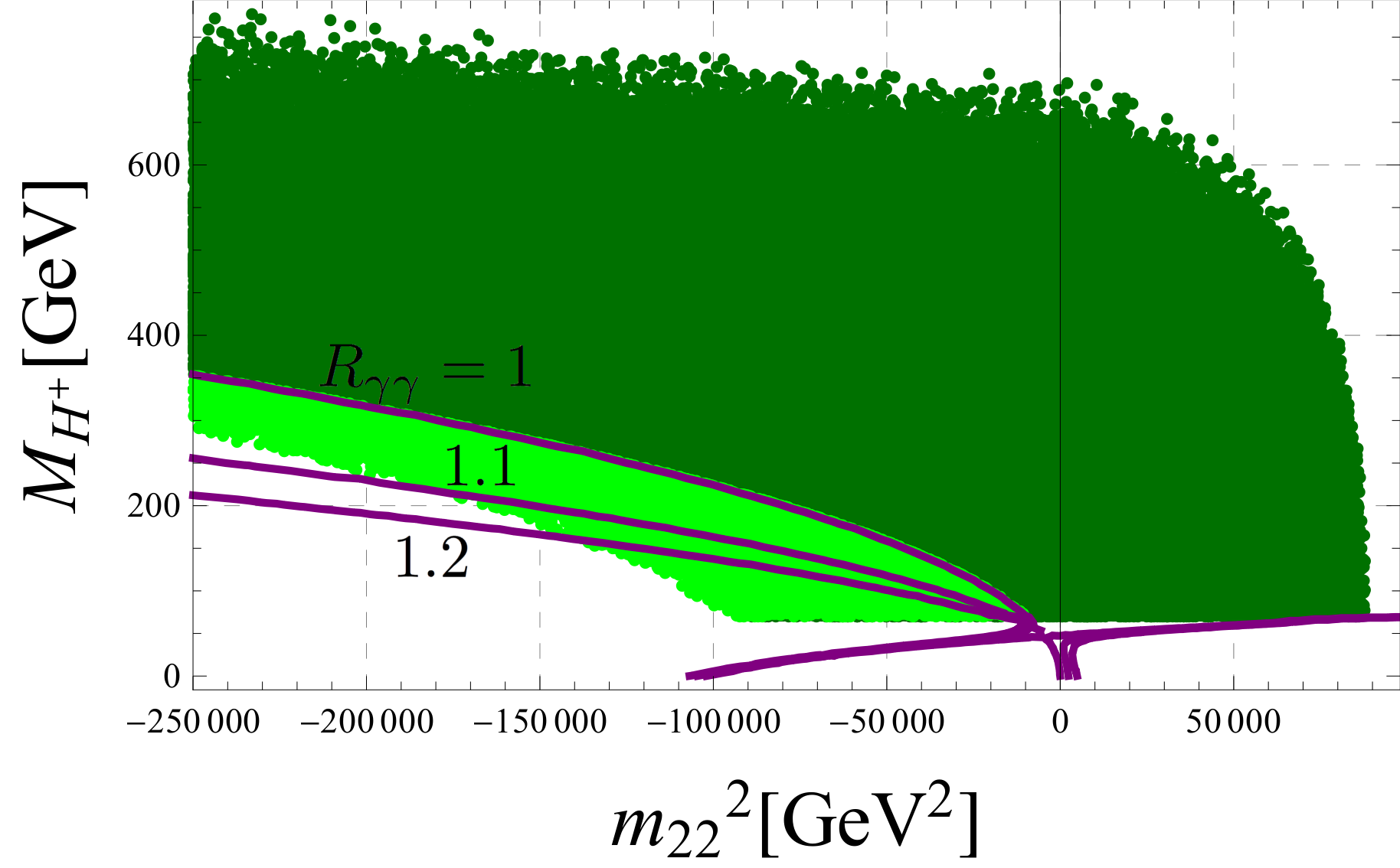}
\caption{Left:  $\rg$ vs $\rzg$. Right: The region allowed by theoretical and experimental constraints in the $(m_{22}^2,\ \m)$ plane. Light green (gray) indicates $\rg\geqslant1$, the curves represent constant values of $\rg$.} \label{1}
\end{figure}

$\rg$ versus $\rzg$ is presented in the left panel of Fig.~\ref{1}~\cite{rg}. It can be observed that the correlation between the two rates is positive. If the invisible channels are open both of the rates are strongly damped (lower branch of the curve is a straight line, $\rg\approx\rzg$), if they are closed, the impact of the charged scalar loop is visible (the upper branch of the curve).

\section{Analysis of the two-photon rate}
We have checked~\cite{rg} that $\rg$ cannot be enhanced if the invisible decay channels are open. This means that if $\rg>1$, then $M_H$, $M_A$, $\m$$>62.5\g$, which would exclude very light DM.

In Fig.~\ref{1} (right panel) the region allowed by theoretical and experimental constraints in the $(m_{22}^2,\ \m)$ plane is presented together with the curves indicating constant values of $\rg$. While the region where $\rg\geqslant1$ [light green (gray)] is not constrained, the region where the enhancement is substantial is bounded. For example, for $\rg>1.2$, $\m<154\g$, and $M_H<154\g$ as well. These results combined with the ones described previously and the LEP bound on $\m$ give the following constrains: $62.5\g<M_H\lesssim154\g$, $70\g<\m\lesssim154\g$~\cite{rg}. The requirement $\rg>1.2$ constrains also the couplings $hHH\sim\lambda_{345}$ and $hAA\sim\lambda_{345}^-$: $-1.45\lesssim\lambda_3, \lambda_{345}\lesssim-0.18$.

\begin{figure}[hb]
\centering
\includegraphics[width=.38\textwidth]{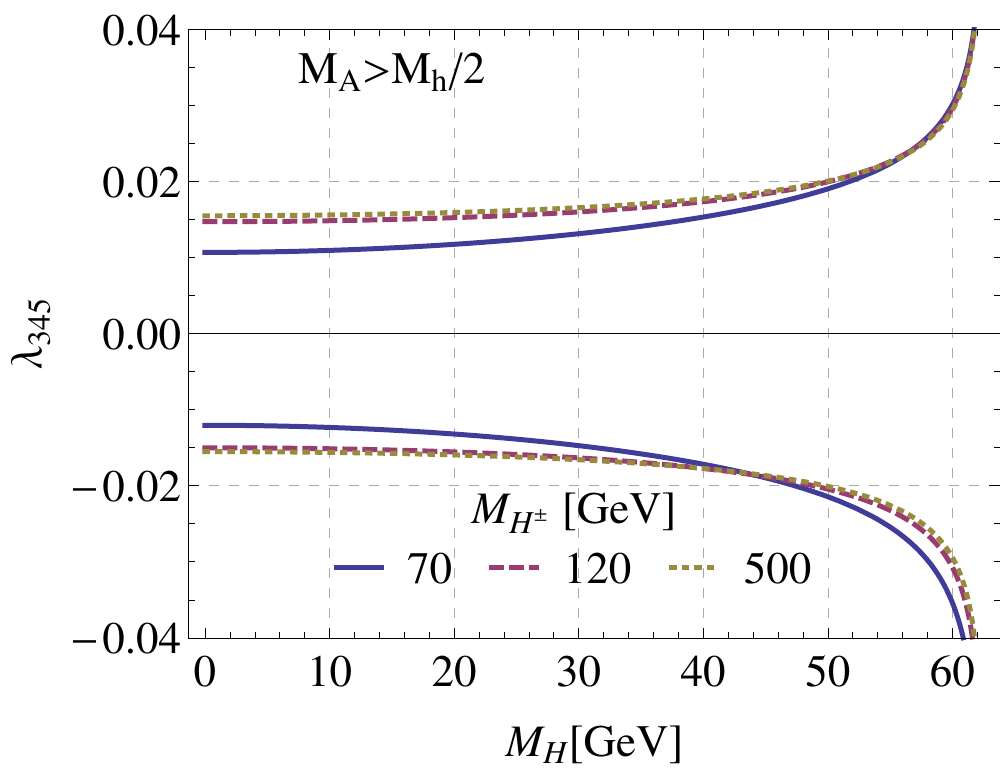}\hspace{1cm}
\includegraphics[width=.36\textwidth]{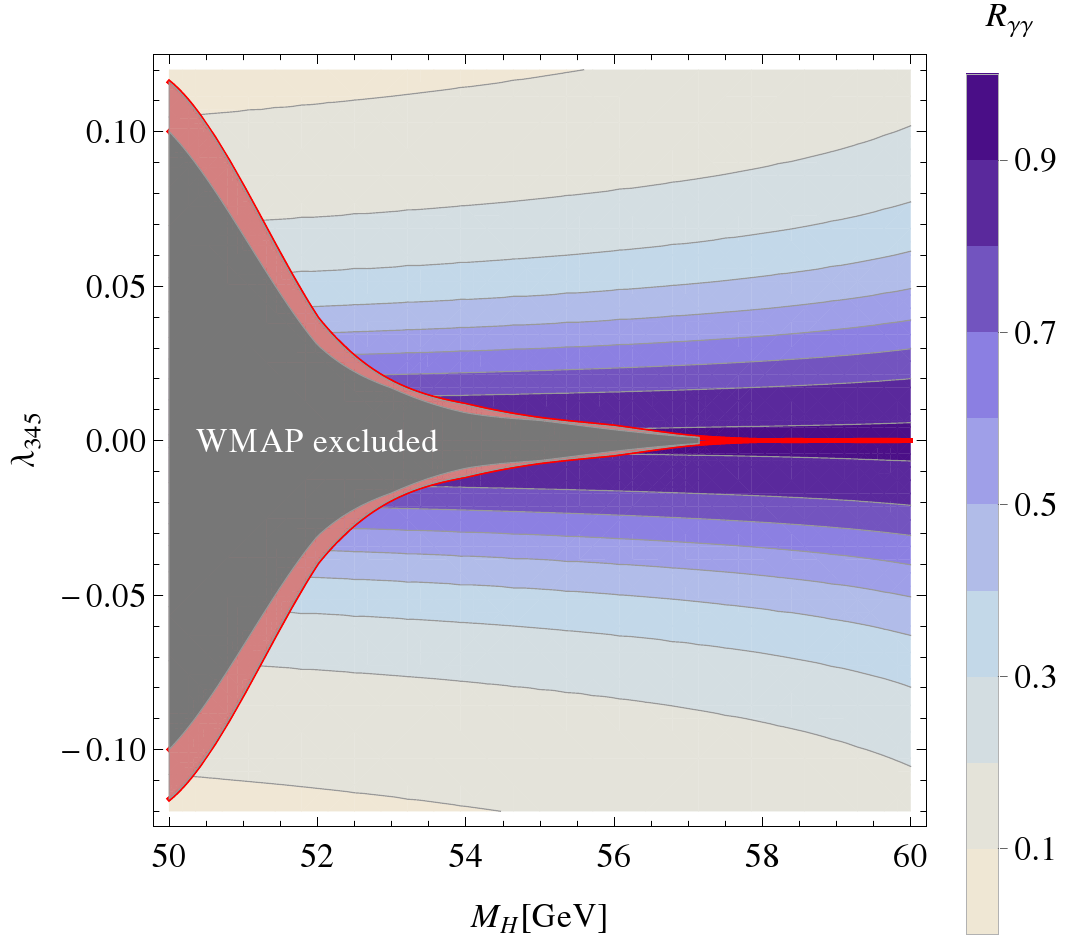}
\caption{Left: Upper and lower bounds on $\lambda_{345}$ following from the requirement $\rg>0.7$ as functions of $M_H$. Right: Comparison of WMAP constraints (red band indicates DM relic density within 3$\sigma$ limits) with the values of $\rg$. The figures come from Ref.~\cite{my-jhep}\label{2}}
\end{figure}

Even if $\rg<1$, setting a lower bound on $\rg$ yields upper and lower bounds on $\lambda_{345}$~\cite{my-jhep}, which are presented in Fig.~\ref{2} (left panel) as functions of $M_H$ (for $M_H<M_h/2$ and $M_A>M_h/2$). We have checked that these constraints are stronger than those following from current measurements of invisible branching ratios of the Higgs boson at the LHC (Br$(h\to\mathrm{inv})<65\%$) and stronger than the XENON100 bounds as well.
In the right panel of Fig.~\ref{2} the $3\sigma$ WMAP limits are superposed on a map showing the values of $\rg$. One can notice that the assumption $\rg>0.7$ is consistent with the WMAP bounds only for $M_H>53\g$, which excludes light DM candidates.
\section{Summary}
Study of the $h\to \gamma\gamma$ and $h\to Z\gamma$ channels can provide important information about the IDM, especially when combined with the WMAP measurements. Requiring $\rg>0.7$ excludes light DM, whereas if $\rg>1.2$, then $62.5\g<M_H\lesssim154\g$, $70\g<\m\lesssim154\g$.
\acknowledgments
I would like to thank M.~Krawczyk, D.~Soko\l owska and P.~Swaczyna for the fruitful collaboration that lead to the results presented in this paper. This work was supported in part by the grant NCN OPUS 2012/05/B/ST2/03306 (2012-2016).


\begin{thebibliography}{99}
\bibitem{idm} N. G. Deshpande, E. Ma, \textit{Pattern of Symmetry Breaking with Two Higgs Doublets}, Phys.Rev. D18 (1978) 2574; R.~Barbieri, L.~J.~Hall, and V.~S.~Rychkov, \textit{Improved naturalness with a heavy higgs: An alternative road to lhc physics}, Phys. Rev. D74 (2006) 015007,  [hep-ph/0603188].
\bibitem{dm} 
D.~Soko\l owska, (2011), \textit{DM Data and Constraints on Quartic Couplings in IDM}, 1107.1991 [hep-ph].
\bibitem{rg}  B.~\'{S}wiezewska, M.~Krawczyk, \textit{Diphoton rate in the inert doublet model}, Phys. Rev. D 88 (2013) 035019, [1212.4100 [hep-ph]]; 
\bibitem{my-jhep} M.~Krawczyk, D.~Soko\l owska, P.~Swaczyna, B.~\'{S}wiezewska, \textit{Constraining Inert Dark Matter by $R_{\gamma\gamma}$ and WMAP data}, JHEP09(2013)055, [1305.6266 [hep-ph]].
\end{thebibliography}
\end{document}